\documentclass[onecolumn,superscriptaddress,noshowpacs,aps,prl]{revtex4}

\usepackage{graphicx}
\usepackage{dcolumn}
\usepackage{multirow}
\usepackage{bm}
\usepackage{times}
\usepackage{color}
\usepackage{amstext}
\usepackage{amsmath}
\usepackage{amsfonts}
\usepackage{ocg,dashbox,color}%
\usepackage[bookmarks=true,colorlinks=false]{hyperref}
\definecolor{lightblue}{rgb}{0.6,0.9,1}

\newcommand{\ToggleLayer}[2]{%
  \leavevmode%
  \pdfstartlink user {
    /Subtype /Link
    /Border [0 0 0]%
    /A <<
      /S/JavaScript
      /JS (
         var aOCGs = this.getOCGs();
         for(var i=0; aOCGs && i<aOCGs.length;i++)
         {
         if(aOCGs[i].name == "#1")
         {aOCGs[i].state = 1} else {aOCGs[i].state = 0};
         }
      )
    >>
  }#2%
  \pdfendlink%
}
\begin{document}

\title{Superconductivity-induced optical anomaly in an iron arsenide}

\author{A. Charnukha}
\email[Correspondence should be addressed to A.~C. and A.~V.~B.: ]{a.charnukha@fkf.mpg.de, a.boris@fkf.mpg.de}
\author{P. Popovich}
\author{Y. Matiks}
\author{D. L. Sun}
\author{C. T. Lin}
\author{A. N. Yaresko}
\author{B. Keimer}
\affiliation{Max Planck Institute for Solid State Research, 70569 Stuttgart, Germany}
\author{A. V. Boris}
\affiliation{Max Planck Institute for Solid State Research, 70569 Stuttgart, Germany}
\affiliation{Department of Physics, Loughborough University, Loughborough, LE11 3TU, United Kingdom}

\begin{abstract}
One of the central tenets of conventional theories of superconductivity, including most models proposed for the recently discovered iron-pnictide superconductors, is the notion that only electronic excitations with energies comparable to the superconducting energy gap are affected by the transition. Here we report the results of a comprehensive spectroscopic ellipsometry study of a high-quality crystal of superconducting $\textrm{Ba}_{0.68}\textrm{K}_{0.32}\textrm{Fe}_2\textrm{As}_2$ that challenges this notion. We observe a superconductivity-induced suppression of an absorption band at an energy of $2.5\ \textrm{eV}$, two orders of magnitude above the superconducting gap energy $2\Delta\sim 20\ \textrm{meV}$. Based on density-functional calculations, this band can be assigned to transitions from As-p to Fe-d orbitals crossing the Fermi surface. We identify a related effect at the spin-density-wave transition in parent compounds of the 122 family. This suggests that As-p states deep below the Fermi level contribute to the formation of the superconducting and spin-density-wave states in the iron arsenides.
\end{abstract}

\maketitle

The standard Bardeen-Cooper-Schrieffer (BCS) theory of superconductivity based solely on an effective attractive interaction between electrons mediated by phonons does not provide a satisfactory explanation of the properties of strongly-correlated high-temperature superconductors. Theoretical proposals going back many years suggest that electronic excitations might enhance this interaction and thus contribute to the formation of the superconducting condensate~\cite{PhysRevB.7.1020,0953-8984-16-35-003,PhysRevB.39.11515}.
These proposals appeared to gain some ground with the observation of superconductivity-induced transfer of the optical spectral weight in the cuprate high-temperature superconductors which involves a high-energy scale extending to the visible range of the spectrum~\cite{RevModPhys.77.721}. In spite of numerous studies (for a comprehensive list of references see ref.~\onlinecite{PhysRevB.81.245111}) no modification of interband optical transitions in the superconducting state has been directly identified in the cuprates. Instead, the observed superconductivity-induced anomalies in the optical response of highly conducting $\rm CuO_2$ planes were found to be confined to the energy range corresponding to transitions within the conduction band below the plasma edge. These changes are dominated by a narrowing of the broad Drude peak caused by superconductivity-induced modification of the scattering rate~\cite{PhysRevB.53.6734,A.V.Boris04302004,PhysRevB.72.144503}. Minute redistribution of the spectral weight between the conducting and high-energy Hubbard bands generated by Coulomb correlations may also play a role~\cite{PhysRevLett.95.097002,PhysRevB.76.104509}.

Current research on the recently discovered iron-pnictide superconductors~\cite{Mazin_NatureInsights_2010} suggests that electronic correlations are weaker than those in the cuprates. Unlike in cuprates, the Fermi surface has been reliably determined over the entire phase diagram and shows good agreement with density functional calculations. The superconducting state of the iron pnictides appears to fit well into a BCS framework in which phonons, which in these compounds interact only weakly with electrons~\cite{boeri:026403}, are replaced by spin fluctuations~\cite{PhysRevLett.101.057003}. The ellipsometric data we present here are consistent with the hypothesis that electronic correlations result in only a modest renormalization of the electronic states. However, superconductivity-induced optical anomalies involve modification of an absorption band peaked at an energy of $2.5\ \textrm{eV}$, two orders of magnitude larger than the superconducting gap $2\Delta\approx20\ \textrm{meV}$. In contrast to cuprate superconductors, this high-energy anomaly has a regular Lorentzian shape in both the real and imaginary parts of the dielectric function and is confined to energies well above the plasma edge $\hbar\omega_{\mathrm{pl}}\approx1.5\ \textrm{eV}$. It can be explained as a consequence of non-conservation of the total number of unoccupied states involved in the corresponding optical transitions due to the opening of the superconducting gap. This implies that unconventional interactions beyond the BCS framework must be considered in models of the superconducting pairing mechanism.
\begin{figure*}[htbp]
\includegraphics[width=0.8 \textwidth]{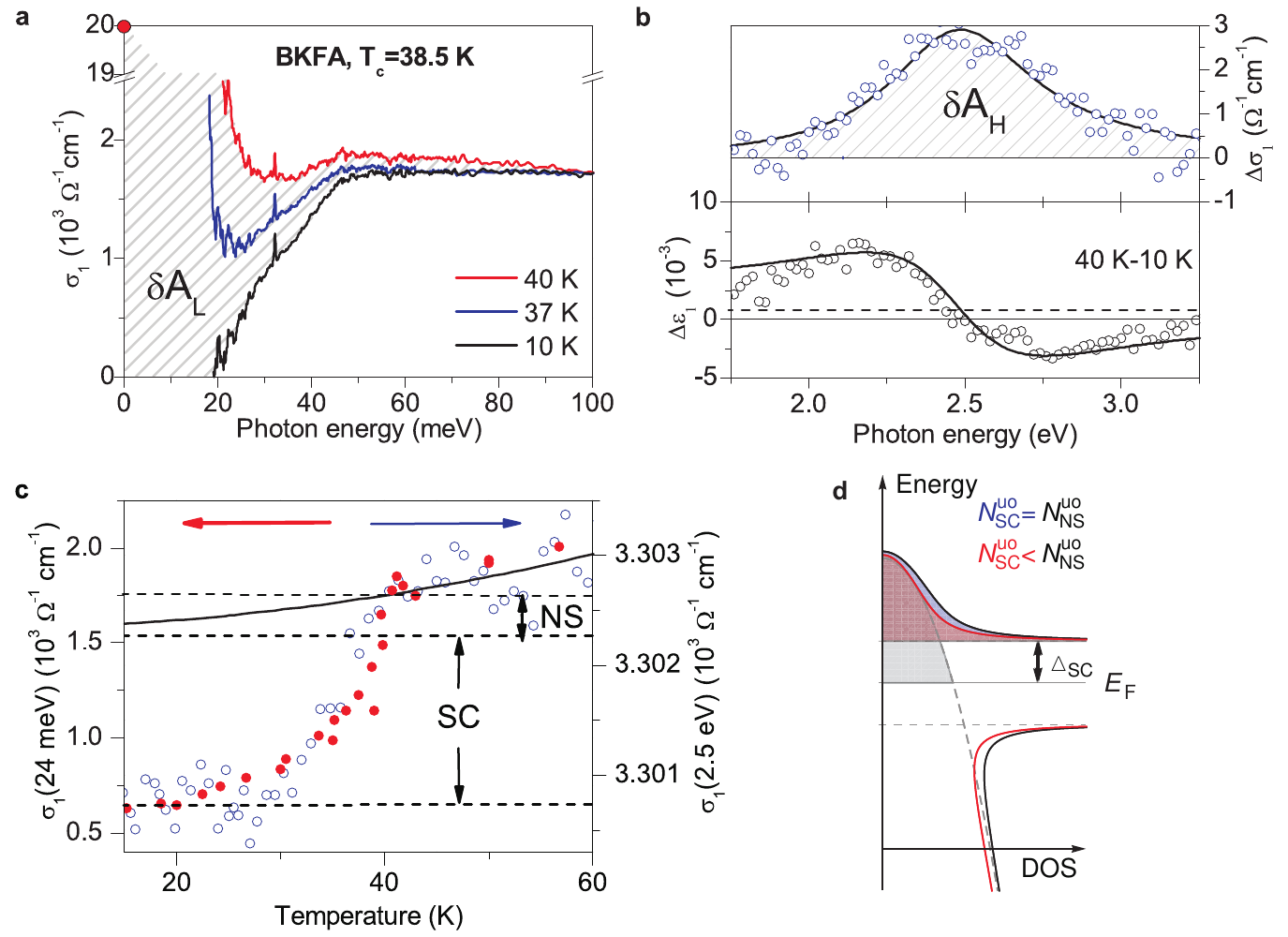}
\caption{\label{fig:scinducedcolouring}\textbf{a},~Real part of the far-infrared optical conductivity of $\textrm{Ba}_{0.68}\textrm{K}_{0.32}\textrm{Fe}_2\textrm{As}_2$ and the missing area. \textbf{b},~Difference spectra of the real part of the optical conductivity~(top panel) and dielectric function~(bottom panel) between $40$ and $10\ \rm{K}$, with a small background shift (horizontal dashed line) detected by temperature modulation measurements~(see Supplementary Information). Lorentzian fit to both spectra (black solid lines). \textbf{c},~Temperature scan at $2.5\ \rm{eV}$. Contribution of the normal-state dynamics (solid line) was estimated to determine the magnitude of the SC-induced jump. \textbf{d},~Density of states in the normal (gray line), conventional superconducting state (black line), and an unconventional state with a depletion of unoccupied states (red line). Filled areas of respective colors represent total number of unoccupied states.}
\end{figure*}

The measurements were carried out on a single crystal of $\textrm{Ba}_{1-x}\textrm{K}_{x}\textrm{Fe}_2\textrm{As}_2$ (BKFA) with $x = 0.32$ and superconducting $T_{\mathrm{c}}=38.5\ \textrm{K}$. Specific-heat measurements on the same sample confirm its high purity and the absence of secondary electronic phases ~\cite{PhysRevLett.105.027003}. We performed direct ellipsometric measurements of the in-plane complex dielectric function $\varepsilon(\omega)=\varepsilon_1(\omega)+i\varepsilon_2(\omega)=1+4\pi i\sigma(\omega)/\omega$ over a range of photon energies extending from the far infrared ($\hbar\omega=12\ \textrm{meV}$) to the ultraviolet ($\hbar\omega=6.5\ \textrm{eV}$) with subsequent Kramers-Kronig consistency analysis~(see Supplementary Information). The far-infrared optical conductivity, is dominated by the opening of a superconducting gap of magnitude $2\Delta \approx20\ \textrm{meV}$ below $T_{\mathrm{c}}$ (Fig.~\ref{fig:scinducedcolouring}a), in accordance with previous studies of optimally doped BKFA~\cite{PhysRevLett.101.107004}. The low-energy missing area in the optical conductivity spectrum below $T_{\mathrm{c}}$, $\delta A_{\mathrm{L}} =\int_{0^+}^{10\Delta}(\sigma_1^{\mathrm{40K}}(\omega)-\sigma_1^{\mathrm{10K}}(\omega))d\omega$, is contained within $10\Delta$ and amounts to $\omega_{\mathrm{pl}}^{\mathrm{sc}}=\sqrt{8\delta A_{\mathrm{L}}}=0.9\ \textrm{eV}$, equivalent to a London penetration depth of $\lambda_{\mathrm{p}}=2200$~\AA. The fraction of the missing area below $12\ \textrm{meV}$ not accessible to the experiment was accurately quantified from the requirement of Kramers-Kronig consistency of the independently measured real and imaginary parts of the dielectric function.

Careful examination of the visible range uncovered superconductivity-induced suppression of an absorption band at $2.5\ \textrm{eV}$. Figure~\ref{fig:scinducedcolouring}b shows difference spectra between $40\ \textrm{and}\ 10\ \textrm{K}$ of the real parts of the optical conductivity and dielectric function. The suppressed band has a Lorentzian lineshape and appears abruptly across the superconducting transition, consistently in both $\Delta\sigma_1$ and $\Delta\varepsilon_1$, as shown in Fig.~\ref{fig:scinducedcolouring}c for $\sigma_1(2.5\ \textrm{eV})$. The temperature dependence of the suppression (blue open circles) coincides with that of the far-infrared optical conductivity due to the opening of the superconducting gap (red filled circles). Thus the onset of superconductivity not only modifies the low-energy quasiparticle response, but also affects the overall electronic structure including interband transitions in the visible range of the spectrum. Since the spectral-weight (SW) loss $\delta A_{\mathrm{H}}$ is not balanced in the vicinity of the absorption band (Fig.~\ref{fig:scinducedcolouring}b), our data indicate a SW transfer over a wide energy range. We note that superconductivity-induced modification of the lattice parameters only results in a minute volume change of $\Delta V/V\approx 5 \times 10^{-7}$, which is insufficient to explain the optical anomaly~\cite{privatecomm_Meingast}. A Kramers-Kronig consistency analysis could not be carried out with sufficient accuracy to show whether or not the SW liberated from the absorption band at the superconducting transition contributes to the response of the superconducting condensate at zero energy. We did, however, detect a minute rise of the background level of $\varepsilon_1(\textrm{1.5-3.5 eV})$ below $T_{\mathrm{c}}$, which according to the Kramers-Kronig relation implies that the SW is transferred to energies below $1.5\ \textrm{eV}$. This effect was identified from a simultaneous fit of $\Delta\sigma_1(\omega)$ and $\Delta\varepsilon_1(\omega)$~(horizontal dashed line in the bottom panel of Fig.~\ref{fig:scinducedcolouring}b). Further accurate temperature modulation measurements of $\sigma_1$ and $\varepsilon_1$ at characteristic energies confirmed a background increase of $\Delta\varepsilon_1(2.5\ \textrm{eV})=(8\pm4)\ 10^{-4}$~(see Supplementary Information).

Since the spectral weight $\delta A_{\mathrm{H}}$ liberated from the absorption band upon cooling below $T_{\mathrm{c}}$ comprises only $\sim 0.5\%$ of the total spectral weight $\delta A_{\mathrm{L}}$ of the superconducting condensate, its contribution to the low-energy charge dynamics might be considered negligible. However, assuming that this additional high-energy SW contributes to the itinerant-carrier response below $T_{\mathrm{c}}$, a simple estimate in the framework of the tight-binding nearest-neighbour approximation~\cite{PhysRevB.16.2437,Hirsch1992305} shows that this would lead to a reduction of electronic kinetic energy of $0.60\ \textrm{meV/unit cell}$ in the superconducting state~(see Supplementary Information). This is close to the condensation energy $\Delta F(0)=0.36\ \textrm{meV/unit cell}$ obtained from specific-heat measurements on the same sample~\cite{PhysRevLett.105.027003}. It is thus important to establish the origin of this unusual optical anomaly.
\begin{figure*}[htbp]
\includegraphics[width=\textwidth]{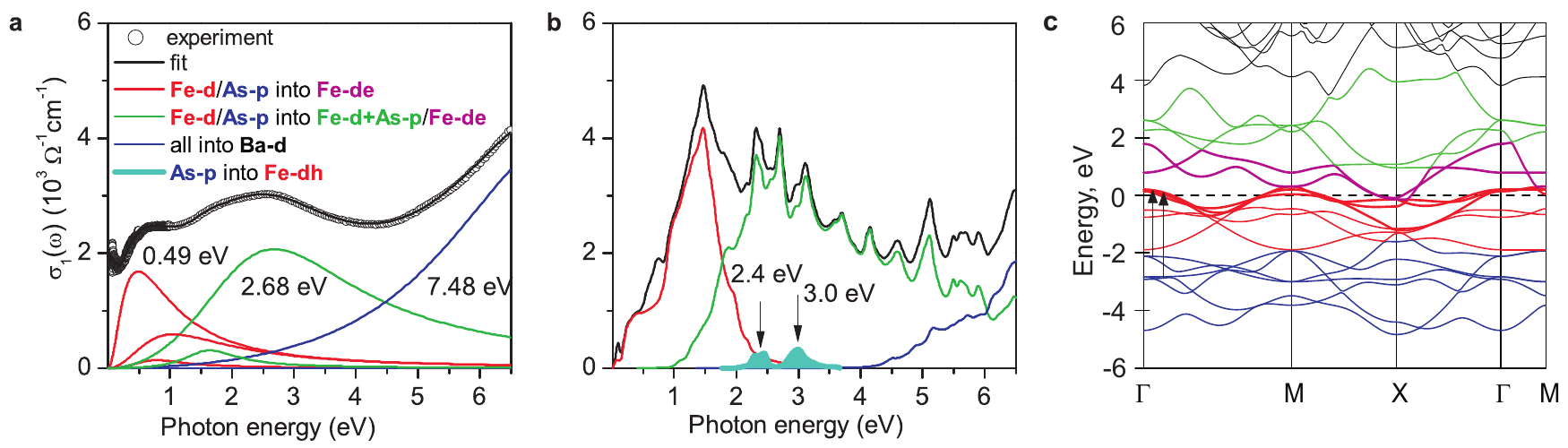}
\caption{\textbf{a},~Real part of the optical conductivity and major interband transitions of $\textrm{Ba}_{0.68}\textrm{K}_{0.32}\textrm{Fe}_2\textrm{As}_2$ determined by a dispersion analysis. \textbf{b},~Corresponding LDA calculation with a breakdown into separate orbital contributions described in the legend~\textbf{a}. \textbf{c},~Band structure from the same LDA calculation. Color coding of the dispersion curves corresponds to the \textit{text} color in legend~\textbf{a}. Superconductivity-suppressed absorption bands (black arrows).}
\label{fig:elstr}
\end{figure*}

We therefore compared our data to the results of ab-initio electronic structure calculations in the framework of the Local Density Approximation (LDA) (Figs.~\ref{fig:elstr}a and~\ref{fig:elstr}b). A dispersion analysis of the the experimental optical conductivity in the range $0.5\ -\ 6.5\ \textrm{eV}$ yielded three major interband transitions in $\textrm{Ba}_{0.68}\textrm{K}_{0.32}\textrm{Fe}_2\textrm{As}_2$. Comparison to the LDA results enabled us to identify the initial and final states of these transitions. The lowest-energy transition is located at about $1\ \textrm{eV}$ (red line) and stems from intraband $\textrm{Fe}-d$ and interband $\textrm{As}-p$ to $\textrm{Fe}-d$ transitions. The major contribution to the optical response in the visible spectral range comes from transitions starting from $\textrm{Fe}-d$ or $\textrm{As}-p$ orbitals into strongly hybridized $\textrm{Fe}-d$ to $\textrm{As}-p$ or $\textrm{Fe}-d$ orbitals (green line). Finally, the UV absorption comes from higher-energy transitions into $\textrm{Ba}-d$ states (blue line).

Although the high-energy electronic structure of BKFA is predicted quite well by the LDA calculations, the experimental quasiparticle response due to transitions within the conduction band (or, given the multiorbital structure of iron pnictides, a narrowly-spaced set of conduction bands) shows a significant deviation. The discrepancy can be quantified by the squared ratio of the band-structure plasma frequency $\omega_{\mathrm{pl}}^{\mathrm{LDA}}=2.7\ \textrm{eV}$ (not included in Fig.~\ref{fig:elstr}b) to its experimental counterpart $\omega_{\mathrm{pl}}^{\mathrm{exp}}=1.5\ \textrm{eV}$, which can be obtained in practice from the residual optical response, after the interband transitions identified using a dispersion analysis have been subtracted. This ratio approximates the quasiparticle effective-mass renormalization factor $m^\star/m_{\mathrm{band}}=\left(\omega^{\mathrm{LDA}}_{\mathrm{pl}}/\omega_{\mathrm{pl}}\right)^2\approx3$. Such an enhancement is consistent with de~Haas-van~Alphen and photoemission experiments on other compounds of the 122 family~\cite{Analytis_BFAP_QO_2010,PhysRevLett.103.076401,PhysRevB.80.024515} and was recently reproduced by combined LDA+DMFT calculations for both 1111 and 122 compounds~\cite{PhysRevB.80.092501}. These calculations do not show evidence of formation of Hubbard bands and thus indicate moderate electron-electron correlations. This explains the good agreement of the LDA optical conductivity above $1.5\ \textrm{eV}$ with the experimental data.

Now we turn to the physical origin of the superconductivity-suppressed absorption band. The same LDA calculation revealed a set of interband transitions centered at $2.5\ \textrm{eV}$, which originate or terminate in states exhibiting hole dispersion and crossing the Fermi level at the $\Gamma$- and $M$-points of the Brillouin zone (Fig.~\ref{fig:elstr}c). We confidently assign these states to $\textrm{Fe-d}_{yz,zx}$ and $\textrm{Fe-d}_{xy}$ orbitals. The other states involved in these transitions belong to $\textrm{As-p}_{x,y}$/$\textrm{Fe-d}_{z^2}$ hybrid orbitals about $2-3\ \textrm{eV}$ below and above the Fermi level, giving rise to a bandwidth of $\Delta E\approx1\ \textrm{eV}$, in remarkable agreement with experiment. Suppression of the absorption band over its full width can be explained by redistribution of the occupation of $\textrm{Fe-d}_{yz,zx}$ and $\textrm{Fe-d}_{xy}$ states under the Fermi level below the superconducting transition. This mechanism is supported by LDA calculations in which the density of states within one superconducting gap energy above the Fermi level was eliminated~(see Supplementary Information) leading to the observed suppression of the optical transitions shown in cyan in Fig.~\ref{fig:elstr}b. 

The required population redistribution is, however, at variance with the conventional theory of superconductivity. In the framework of the standard BCS approach, opening of an energy gap in a single-band superconductor leads to a bending of the quasiparticle dispersion and an expulsion of the density of states in the vicinity of the Fermi surface (gray and blue areas in Fig.~\ref{fig:scinducedcolouring}d)~\cite{Tinkham_superconductivity_1995} and does not lead to population redistribution, i.e. the total number of unoccupied states below the transition is conserved $N^{\mathrm{uo}}_{\mathrm{SC}}=N^{\mathrm{uo}}_{\mathrm{NS}}$ (blue area is equal to the gray area in Fig.~\ref{fig:scinducedcolouring}d). This can only lead to a small corrugation of an optical absorption band on the scale of one superconducting-gap energy superimposed on the overall broad feature without any modification of its spectral weight~\cite{Dobryakov1994309}. The experimentally observed suppression of an absorption band \textit{on the scale of its full width} necessarily requires population imbalance $N^{\mathrm{uo}}_{\mathrm{SC}}<N^{\mathrm{uo}}_{\mathrm{NS}}$ (red area unequal to the gray area in Fig.~\ref{fig:scinducedcolouring}d). This effect can be clearly identified as a consequence of superconductivity because the temperature dependence of the suppression mimics that of the optical conductivity in the FIR region due to the opening of the superconducting gap, as shown in Fig.~\ref{fig:scinducedcolouring}c.

All of the iron-pnictide superconductors are known to have multiple superconducting gaps~\cite{Mazin_NatureInsights_2010} and theoretical work indicates a dominant contribution of electron pairing between different bands to the formation of the superconducting state~\cite{PhysRevLett.101.057003}. Redistribution of the occupation of the different bands below $T_{\mathrm{c}}$ could explain the optical anomaly we observed. It requires a lowering of the material's chemical potential in the superconducting state. However, even a generalization of the standard BCS theory to the multiband case~\cite{PhysRevLett.3.552} does not take into account this effect. Therefore, self-consistent treatment of a variable chemical potential at the superconducting transition is needed. In the presence of large Fe-As bond polarizability~\cite{PhysRevB.79.214507} it can potentially enhance superconductivity in iron pnictides.
\begin{figure*}[htbp]
\includegraphics[width=\textwidth]{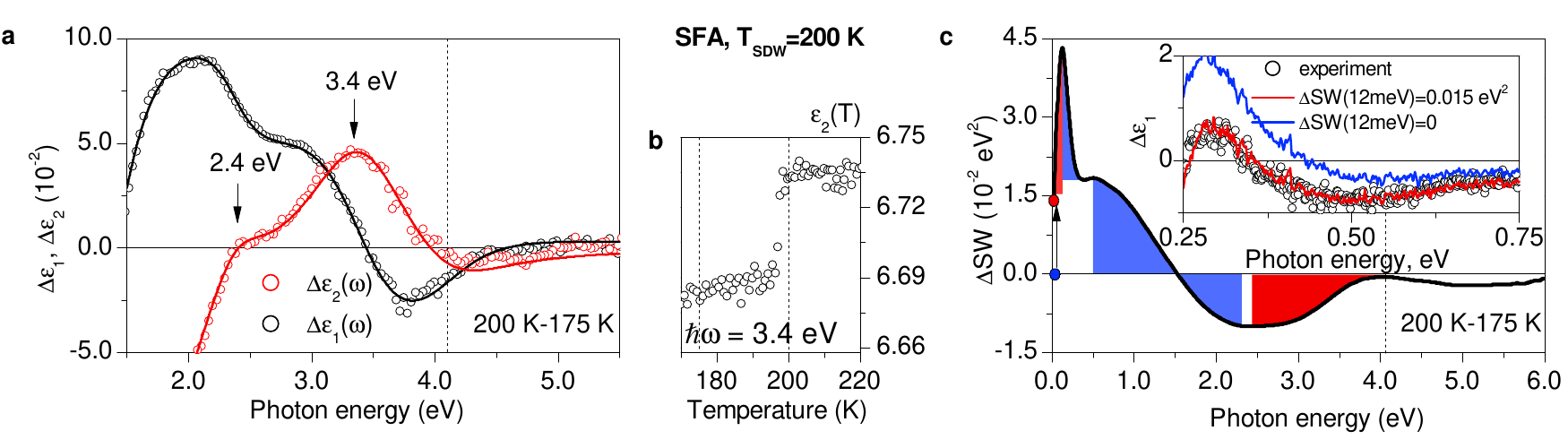}
\caption{\textbf{a},~Real and imaginary parts of the dielectric function of $\rm{SrFe}_2\rm{As}_2$. Lorentzian fit to both (solid lines). \textbf{b},~Temperature scan at $3.4\ \textrm{eV}$. \textbf{c},~Spectral weight~(SW) redistribution between $200\ \textrm{K}$ and $175\ \textrm{K}$. Extrapolation-dependent SW before (blue) and after (red) a Kramers-Kronig (KK) consistency check. Blue and red filled areas represent regions of SW gain and loss, respectively, in the magnetic versus normal state. inset, Real part of the dielectric function and KK transformations of the real part of the optical conductivity (solid lines, colors match filled circles).}
\label{fig:sfa}
\end{figure*}
We have further explored the validity of this scenario by repeating our ellipsometric measurements on parent compounds of the 122 family of iron arsenide superconductors. Since the spin-density-wave (SDW) instability exhibited by these compounds is also believed to be induced by nesting of electronic states on different electronic bands, we expect a similar optical anomaly at the SDW transition as the one we observed in the superconductor. The magnitude of the anomaly is expected to be larger than the one in the superconductor, because the SDW transition occurs at a higher temperature and generates a larger energy gap. In $\textrm{SrFe}_2\textrm{As}_2$ (SFA), we indeed find a strong reduction of optical absorption upon cooling below $T_{\mathrm{SDW}} = 200\ \textrm{K}$. The difference spectra of $\varepsilon_1(\omega)$ and $\varepsilon_2(\omega)$  between $200$ and $175\ \textrm{K}$ show a double-peak structure with maxima at $2.4$ and~$3.4\ \textrm{eV}$ (Fig.~\ref{fig:sfa}a). A temperature scan across $T_{\mathrm{SDW}}$ at the frequency of the second peak (Fig.~\ref{fig:sfa}b) further confirms that this effect is induced by SDW formation. A direct comparison with the superconducting compound is complicated by a pronounced modification of the electronic structure due to the coincident magnetic and structural transitions. Nevertheless, certain information can be gained from the critical behavior of the in-plane spectral weight $\Delta SW(\Omega)=\int_0^{\Omega}\Delta\sigma_1(\omega)d\omega$. Figure~\ref{fig:sfa}c shows difference spectral weight of $\textrm{SrFe}_2\textrm{As}_2$ between $200\ \textrm{K}$ and $175\ \textrm{K}$ as a function of the upper integration limit $\Omega$. The change of the SW in the extrapolation region below $12\ \textrm{meV}$ was accurately determined via a Kramers-Kronig consistency analysis~(see Supplementary Information), as illustrated in the inset of Fig.~\ref{fig:sfa}c. With $\Delta SW(12\ \textrm{meV})=0$ across the transition (blue filled circle) the Kramers-Kronig transformation of the $\Delta\varepsilon_2(\omega)$ (blue line) deviates significantly from experimentally measured $\Delta\varepsilon_1(\omega)$. Gradually increasing this SW brings them closer together until they finally coincide (red line) thus fixing $\Delta SW(12\ \textrm{meV})=0.015\ \textrm{eV}^2$ (red filled circle). The higher-energy redistribution of the SW is broken down in Fig.~\ref{fig:sfa}c into regions of SW gain (blue areas) and loss (red areas) in the SDW with respect to paramagnetic state. The SW lost due to the opening of the SDW gap~\cite{PhysRevLett.101.257005} (the first red region) is partly transferred to the electronic excitations across the gap (the first blue region) and fully recovered by $1.5\ \textrm{eV}$. These processes are then followed by higher-energy redistribution in the region from $1.5$ to $4.0\ \textrm{eV}$ involving the SW of the suppressed bands. It appears unlikely that such high-energy SW transfer could result from modification of the electronic structure due to a magnetic transition because effects of electronic reconstruction at the SDW transition are limited by $1.5\ \textrm{eV}$. A modification of the matrix elements at the structural transitions of sufficient strength cannot account for the observed suppression, because this would be accompanied by an even larger effect at higher energies, clearly absent in Fig.~\ref{fig:sfa}c. A redistribution of charge carriers between the SDW-coupled bands analogous to that in the superconducting compound provides a more natural explanation. The same physical reasons might explain the orbital polarization that breaks the degeneracy of $\textrm{Fe-d}_{xz}$ and $\textrm{Fe-d}_{yz}$ orbitals recently observed in the Ba-based parent of the same family by photoemission spectroscopy~\cite{PhysRevLett.104.057002}.

Interactions of electrons in different energy bands at the Fermi level may provide a common framework for an explanation of the optical anomalies in the spin-density-wave and superconducting compounds. It is important to note that these anomalies affect only a small fraction of the interband transitions, which involve initial states of As $p$-orbital character deep below the Fermi level. This indicates that these orbitals significantly influence electronic instabilities in the iron arsenides, possibly due to the high polarizability of As-Fe bonds. Our study points to optical spectral-weight transfer from high energies to below $1.5\ \textrm{eV}$ induced by collective electronic instabilities. In the superconductor, it occurs at energies two orders of magnitude larger than the superconducting gap energy, suggesting that electronic pairing mechanisms contribute to the formation of the superconducting condensate.

\section{Acknowledgements}This project was supported by the German Science Foundation under grant BO~3537/1-1 within SPP 1458. We gratefully acknowledge Y.-L. Mathis for support at the infrared beamline of the synchrotron facility ANKA at the Karlsruhe Institute of Technology and V.~Khanna for taking part in some of the measurements. We also thank O.~V.~Dolgov, L.~Boeri, F.~V.~Kusmartsev, A.~S.~Alexandrov, I.~I.~Mazin, P.~B.~Littlewood for fruitful discussions.

\newpage
\section{Supplementary information}
\setcounter{figure}{0}
\renewcommand\thefigure{S\arabic{figure}}
\renewcommand{\cite}[1]{[S\citenum{#1}]}
\renewcommand{\bibnumfmt}[1]{[S#1]}
\subsection{Experimental}
The $\textrm{Ba}_{0.68}\textrm{K}_{0.32}\textrm{Fe}_2\textrm{As}_2$ single crystal was grown in zirconia crucibles sealed in quartz ampoules under argon atmosphere~\cite{SLin_BKFA_growth_2008}. Its chemical composition was determined by energy-dispersive X-ray spectrometry. The quality of the sample and absence of phase separation was confirmed by an estimated from the residual low-temperature electronic specific heat non-superconducting fraction of less than $2.4\%$~\cite{SPhysRevLett.105.027003}. From DC resistivity, magnetization and specific-heat measurement we obtained $T_{\textrm{c}}=38.5\pm0.2\ \textrm{K}$. The sample surface was cleaved prior to every measurement.

The experimental setup comprises three ellipsometers to cover the spectral range of $12\ \mathrm{meV} - 6.5\ \mathrm{eV}$. For the range $12\ \mathrm{meV} - 1\ \mathrm{eV}$ we used a home-built ellipsometer attached to a standard Fast-Fourier-Transform Bruker 66v/S FTIR interferometer. The FIR measurements were performed at the infrared beamline of the ANKA synchrotron light source at Karlsruhe Institute of Technology, Germany. For the MIR measurements we used the conventional glow-bar light source of a Bruker 66v/S FTIR. Finally, high-energy spectra $0.7\ \mathrm{eV} - 6.5\ \mathrm{eV}$ were measured with a Woollam VASE (Variable Angle Spectroscopic Ellipsometer) ellipsometer equipped with a UHV cold-finger cryostat operated at $<5\times10^{-9}\ \mathrm{mbar}$ chamber pressure.

The inherent capacity of Woollam VASE ellipsometers to measure relative changes of the dielectric function on the order of $10^{-2}$ was boosted to an unprecedented level of $10^{-4}$ using temperature-modulation measurements (see Supplementary Information). The sample temperature was changed between $20$ to $40\ \textrm{K}$ with a period of $1800\ \textrm{s}$ and later averaged over 24 periods to achive sufficient accuracy required to confirm a minute background increase of $\Delta\varepsilon_1(2.47\ \textrm{eV})=(8\pm4)\ 10^{-4}$.

One of the strong advantages of spectroscopic ellipsometry over reflectometry is that independently obtained $\Delta\varepsilon_1(\omega)$ and $\Delta\sigma_1(\omega)$ can be used in a Kramers-Kronig consistency check, in which independently obtained spectra $\Delta\varepsilon_1(\omega)$ and $8\wp\int_0^{\infty}\frac{\Delta\sigma_1(x)}{x^2-\omega^2}dx$ must coincide. This additional constraint unique to ellipsometry allows one to determine with high accuracy the spectral weight in the extrapolation region beyond the experimentally accessible spectral range, in our case below $12\ \textrm{meV}$. This drastically reduces the extrapolation uncertainty and renders subsequent data analysis more robust (see Supplementary Information).

The band-structure calculations were performed using a linear-muffin-tin orbital method in the atomic sphere approximation~\cite{SPhysRevB.12.3060} within the LDA starting form the known crystal structure of $\textrm{Ba(Sr)}_{1-x}\textrm{K}_x\textrm{Fe}_2\textrm{As}_2$.
\subsection{Kramers-Kronig consistency check}
Ellipsometry has an advantage of measuring the complex dielectric function $\varepsilon(\omega)=\varepsilon_1(\omega)+i\varepsilon_2(\omega)=1+4\pi i\sigma(\omega)/\omega$ directly without a need for reference measurements and Kramers-Kronig transformation (KKT). The KKT allows for a consistency check of $\varepsilon_1(\omega)$ and $\sigma_1(\omega)$ and implies that $\Delta\varepsilon_1$ at any energy depends on $\Delta\sigma_1(\omega)$ in the whole spectral range including the extrapolation region:\[\Delta\varepsilon_1(\omega)=8\wp\int_0^{\infty}\frac{\Delta\sigma_1(x)}{x^2-\omega^2}dx,\]where $\varepsilon_0$ is the dielectric permittivity, and $\sigma_1(\omega)$ is the real part of optical conductivity. This consistency analysis is rather insensitive to the exact shape of the extrapolation chosen but it does fix the total spectral weight \[\Delta\mathrm{SW}(\omega_0)=\int_0^{\omega_0}\Delta\sigma_1(x)dx,\]where $\omega_0$ is the experimental low-energy cutoff frequency. This procedure is illustrated in Fig.~3~c of the main text. Taking the experimental difference spectrum of $\Delta\sigma_1(\omega)$ with $\Delta\mathrm{SW}(\omega_0)=0$ (blue circle) and carrying out the KKT results in the real part of the dielectric function deviating from the measured data (blue line in the inset). Only by increasing the spectral weight below $\omega_0=12\ \mathrm{meV}$ to $0.015\ \mathrm{eV}^2$ (red circle) does one achieve complete agreement with the experiment (red line in the inset). The exact shape of the extrapolated $\Delta\sigma_1(\omega)$ plays a minor role. The maximum uncertainty introduced by the unknown shape can be calculated as the difference of two extreme configurations: all $\Delta\mathrm{SW}(\omega_0)$ at $\omega=0$ and $\omega=\omega_0$:\begin{eqnarray}\delta\Delta\varepsilon_1^{(1)}(\omega)=8\left[-\frac{\Delta\mathrm{SW}(\omega_0)}{\omega^2}-\frac{\Delta\mathrm{SW}(\omega_0)}{\omega_0^2-\omega^2}\right]\\=8\frac{\Delta\mathrm{SW}(\omega_0)}{\omega^2}\frac{\omega_0^2}{\omega_0^2-\omega^2}\longrightarrow8\frac{\Delta\mathrm{SW}(\omega_0)}{\omega^2}\frac{\omega_0^2}{\omega^2},\end{eqnarray}when $\omega\gg \omega_0$. On the other hand, the accuracy to which the spectral weight is determined at the same energy is given by\[\left|\delta\Delta\varepsilon_1^{(2)}(\omega)\right|=\left|8\int_{0^+}^{\infty}\frac{\delta\Delta\sigma_1(x)}{x^2-\omega^2}dx\right|\leq\left|8\frac{\delta(\Delta\mathrm{SW}(\omega_0))}{\omega^2}\right|.\]The relative effect of the shape change over the magnitude change of the spectral weight in the extrapolation region is then $|\delta\Delta\varepsilon_1^{(1)}(\omega)/\delta\Delta\varepsilon_1^{(2)}(\omega)|\longrightarrow(\omega_0/\omega)^2$, for $\omega\gg\omega_0$. In the present case taking $\omega_0=12\ \mathrm{meV}$ and $\omega=250\ \mathrm{meV}$~(as in the inset of Fig.~3~c) one gets a shape uncertainty fraction of $0.2\%$. Thus the effect is negligible already at rather low frequencies. The same analysis applies for the high-energy extrapolation above $6.5\ \textrm{eV}$. However, complete agreement between $\Delta\varepsilon_1(\omega)$ and $\Delta\varepsilon_2(\omega)$ within the accuracy of the experiment was found up to $6.5\ \textrm{eV}$, therefore no experimentally discernable missing spectral weight is contained at higher energies.
\begin{figure}[ht]
\includegraphics[width=\textwidth]{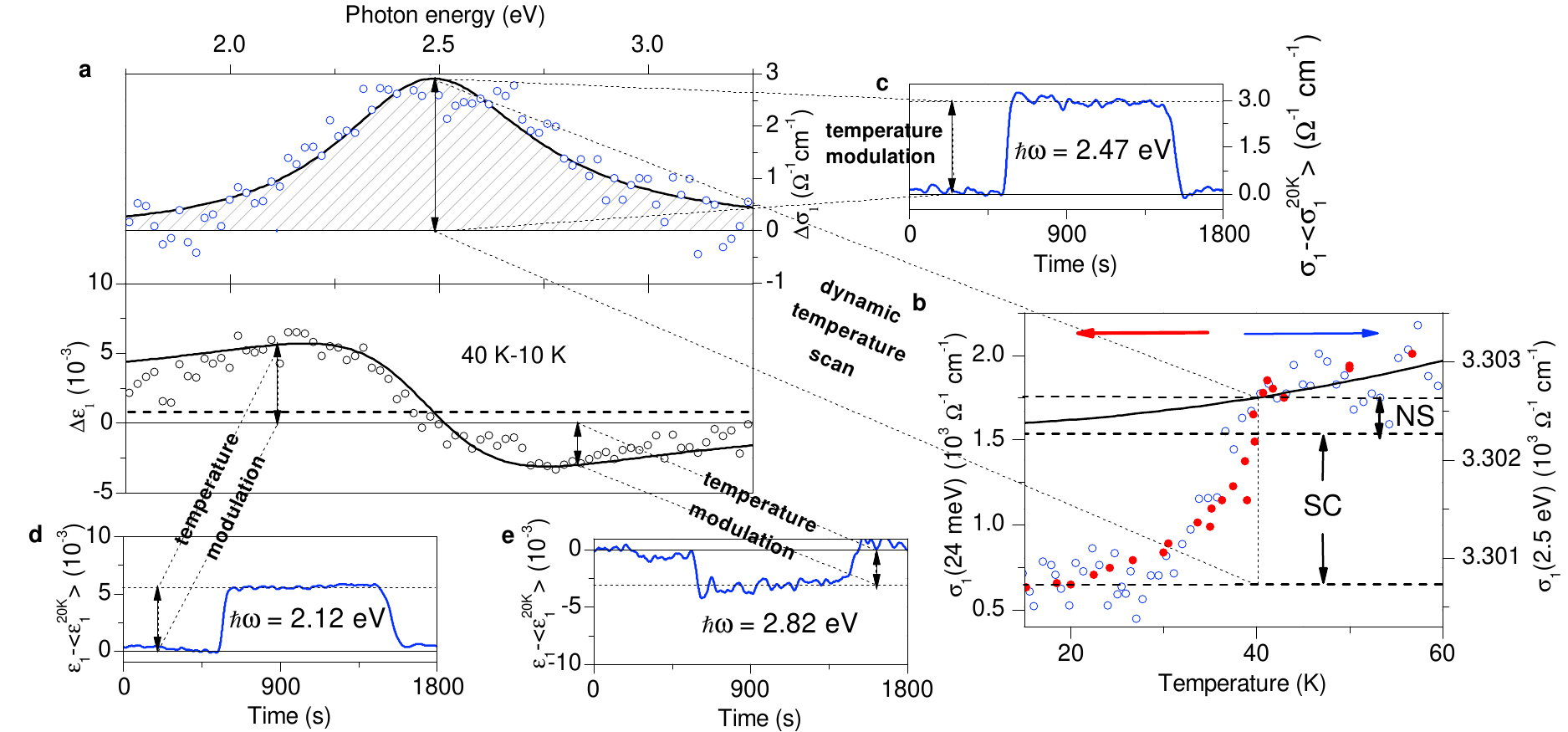}
\caption{\label{fig:bkfatms}\textbf{a}~Difference spectrum of the real part of optical conductivity (top panel) and dielectric function (bottom panel) between $40$ and $10\ \textrm{K}$ (red and black open circles, respectively). Lorentzian fit to both spectra (black solid lines). Small background shift (dashed line in the bottom panel) obtained from the fit and further confirmed by temperature modulation measurements~\textbf{c}-\textbf{e}. \textbf{b}~Temperature scan of the real part of the optical conductivity between $60$ and $15\ \textrm{K}$ at $2.5\ \textrm{eV}$ (blue open circles, right scale) and $24\ \textrm{meV}$ (red filled circles, left scale). \textbf{c}--\textbf{e}~Temperature modulation of the real part of the optical conductivity~(\textbf{c}) and dielectric function~(\textbf{d,e}) between $20\ \textrm{K}$ and $40\ \textrm{K}$ with a period of $1800\ \textrm{sec}$ averaged over $24$~waveforms.}
\end{figure}
\subsection{Temperature-modulation measurements}
The small background shift shown as a dashed line in the bottom panel of Fig.~\ref{fig:bkfatms}a was identified from a simultaneous fit of $\Delta\sigma_1(\omega)$ and $\Delta\varepsilon_1(\omega)$ (upper and bottom panels in Fig.~\ref{fig:bkfatms}a). The superconductivity-induced nature of the suppression of the absorption band as well as of the background shift was confirmed by a dynamic temperature scan between $60$ and $15\ \textrm{K}$ (red open circles in Fig.~\ref{fig:bkfatms}b). Its temperature dependence clearly follows that of the far-infrared optical conductivity due to the opening of the superconducting gap (blue filled circles in Fig.~\ref{fig:bkfatms}b). To estimate the background shift more accurately, temperature modulation measurements of $\sigma_1$ at the resonance photon energy $2.47\ \textrm{eV}$ and $\varepsilon_1$ at off-resonance photon energies of $2.12\ \textrm{eV}$ and $2.82\ \textrm{eV}$ were carried out. In Fig.~\ref{fig:bkfatms}c-e the sample temperature was changed between $20$ to $40\ \textrm{K}$ with a period of $1800\ \textrm{s}$ and later averaged over 24 periods to reduce noise to $\Delta\varepsilon_1=10^{-4}$. This confirms a minute background increase of $\Delta\varepsilon_1(2.47\ \textrm{eV})=(8\pm4)\ 10^{-4}$.
\subsection{LDA calculations: effect of $E_{\mathrm{F}}\pm\Delta$ cuts of the density of states on optical conductivity.}
\begin{figure}[ht]
\includegraphics[width=\textwidth]{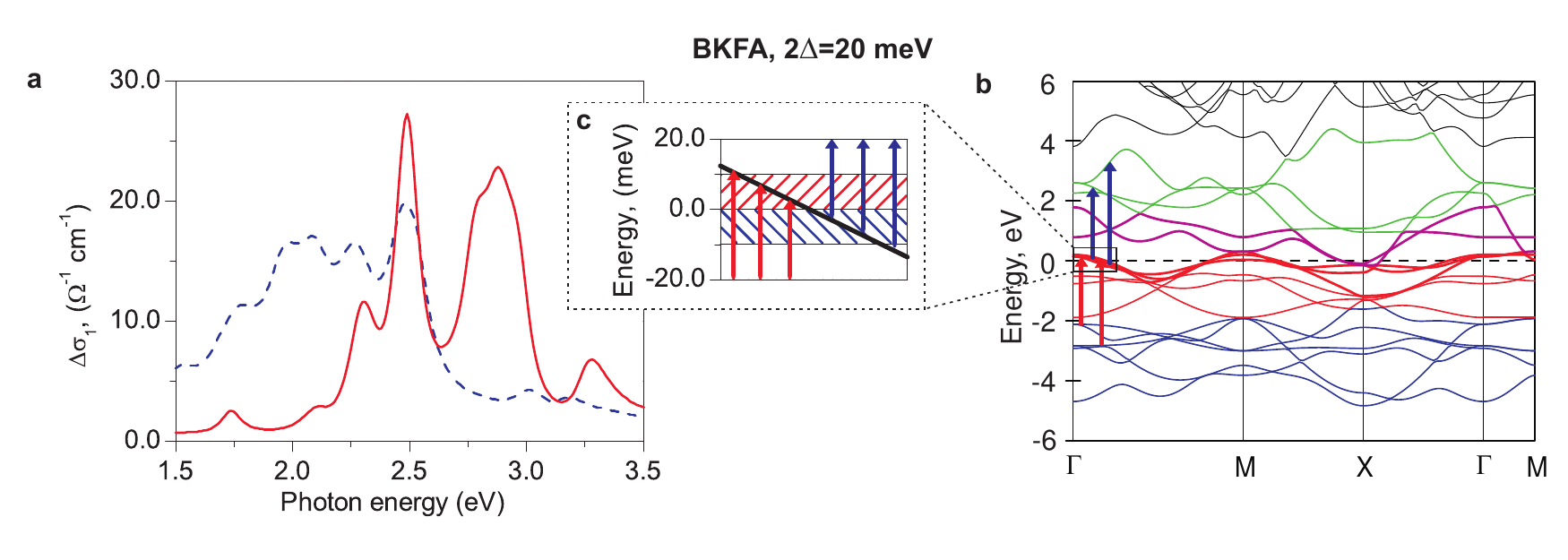}
\caption{\label{fig:bkfaredistribution}\textbf{a}~Difference real part of the optical conductivity between ungapped and gapped regimes for transitions to (red solid line) and from (blue dashed line) the Fermi level. \textbf{b}~Schematic representation of the optical transitions at $\Gamma$ point of the Brillouin zone with a contribution to the optical conductivity shown in~\textbf{a}. \textbf{c}~Population (red) or depletion (blue) of all electronic states within one superconducting gap value $\Delta=10\ \textrm{meV}$. Arrows depict the same optical transitions as in~\textbf{b}.}
\end{figure}

The band-structure calculations were performed using a linear-muffin-tin orbital method in the atomic sphere approximation~\cite{SPhysRevB.12.3060} within the LDA starting form the known crystal structure of $\textrm{Ba(Sr)}_{1-x}\textrm{K}_x\textrm{Fe}_2\textrm{As}_2$~\cite{SPhysRevB.78.020503,S0953-8984-20-45-452201}. The calculations for $\textrm{Ba}_{0.68}\textrm{K}_{0.32}\textrm{Fe}_2\textrm{As}_2$ predict a contribution to the optical conductivity in the visible that results from transitions with final states crossing the Fermi surface, which can experience suppression due to a redistribution of occupied states below the Fermi level across the superconducting transition. Apart from the interband transitions shown as two arrows in Fig.~2~c, a contribution from transitions with initial states crossing the Fermi surface and into higher-lying hybridized As-p and Fe-d states is significant in this energy range. To substantiate that a redistribution of occupied states below the Fermi level within one superconducting gap can account for the observed suppression, optical conductivity in the LDA framework was calculated for the two cases of hole and electron transitions in the vicinity of $\Gamma$- and M-points of the Brillouin zone, i.e. for the hole pockets of the Fermi surface, shown in Fig.~\ref{fig:bkfaredistribution}~a and~\ref{fig:bkfaredistribution}~b, respectively. 
Shown in the Fig.~\ref{fig:bkfaredistribution}~c is difference optical conductivity spectra between gapped and ungapped regime. In the gapped case the transitions in the energy window $\Delta E=\Delta=10\ \textrm{meV}$ below (blue line) and above (red line) the Fermi level are forbidden, to simulate complete depletion (population) of occupied (empty) states. Hole (electron) contribution to the optical conductivity displays suppression only when states above (below) the Fermi level are fully populated (depopulated) to within one superconducting gap. The size of the effect in both cases is an order of magnitude larger than the experimentally detected suppression. Therefore a fractional population redistribution can indeed account for the superconductivity- and SDW-induced anomalies in $\textrm{Ba}_{0.68}\textrm{K}_{0.32}\textrm{Fe}_2\textrm{As}_2$ and $\textrm{SrFe}_2\textrm{As}_2$.
\subsection{Interband optical conductivity in single- and multiband BCS theory}
In the framework of the BCS theory the charge carriers and elementary excitations in the superconducting state differ significantly from those of in the normal state. This is manifested in the modified with respect to free charge carriers dispersion of the excitations of the superconducting condensate - Bogolubov quasiparticles. To obtain this dispersion one can introduce quasiparticle operators diagonalizing the original BCS Hamiltonian~\cite{SPhysRev.108.1175,STinkham_superconductivity_1995} \[\mathcal{H}=\sum_{\mathbf{k}\sigma}\epsilon_{\mathbf{k}}n_{\mathbf{k}\sigma}+\sum_{\mathbf{k}\mathbf{l}}V_{\mathbf{k}\mathbf{l}}c_{\mathbf{k}\uparrow}^\dagger c_{-\mathbf{k}\downarrow}^\dagger c_{-\mathbf{l}\downarrow}^{\mathstrut}c_{\mathbf{l}\uparrow}^{\mathstrut}\]as follows:\begin{eqnarray}
c_{\mathbf{k}\uparrow}^{\mathstrut}&=&u_{\mathbf{k}}^{\mathstrut *}\gamma_{\mathbf{k}0}^{\mathstrut}+v_{\mathbf{k}}^{\mathstrut}\gamma_{\mathbf{k}1}^{\dagger},\nonumber\\
c_{-\mathbf{k}\downarrow}^{\mathstrut}&=&-v_{\mathbf{k}}^{\mathstrut *}\gamma_{\mathbf{k}0}^{\mathstrut}+u_{\mathbf{k}}^{\mathstrut}\gamma_{\mathbf{k}1}^{\dagger}\label{eq:bogolubov}.\end{eqnarray}The complex functions of the k-vector $u_{\mathbf{k}}^{\mathstrut}$ and $v_{\mathbf{k}}^{\mathstrut}$ determine the probability of the pair state comprised of electrons with momenta $\mathbf{k}$ and $-\mathbf{k}$ being empty or occupied, respectively:\begin{eqnarray}\left|u_{\mathbf{k}}^{\mathstrut}\right|^2&=&\frac{1}{2}\left(1+\frac{\xi_{\mathbf{k}}^{\mathstrut}}{E_{\mathbf{k}}^{\mathstrut}}\right),\nonumber\\
\left|v_{\mathbf{k}}^{\mathstrut}\right|^2&=&\frac{1}{2}\left(1-\frac{\xi_{\mathbf{k}}^{\mathstrut}}{E_{\mathbf{k}}^{\mathstrut}}\right),\label{eq:scprobabilities}
\end{eqnarray}where $E_{\mathbf{k}}^{\mathstrut}=\sqrt{\xi_{\mathbf{k}}^{2}+\Delta^2}$ is the Bogolubov quasiparticle's dispersion and $\xi_{\mathbf{k}}^{\mathstrut}=\epsilon_{\mathbf{k}}^{\mathstrut}-\mu$ is the normal-state electron dispersion measured with respect to the chemical potential $\mu$. This probability distributions are inherently smeared around the Fermi level in the ground state. This smearing of the quasiparticle occupation probabilities at $0\ \textrm{K}$ resembles closely that of normal-state particles at $T=T_{\mathrm{c}}$~\cite{STinkham_superconductivity_1995}.

\begin{figure}[t]
\begin{ocg}{normalstate}{1}{0}
\includegraphics[width=\textwidth]{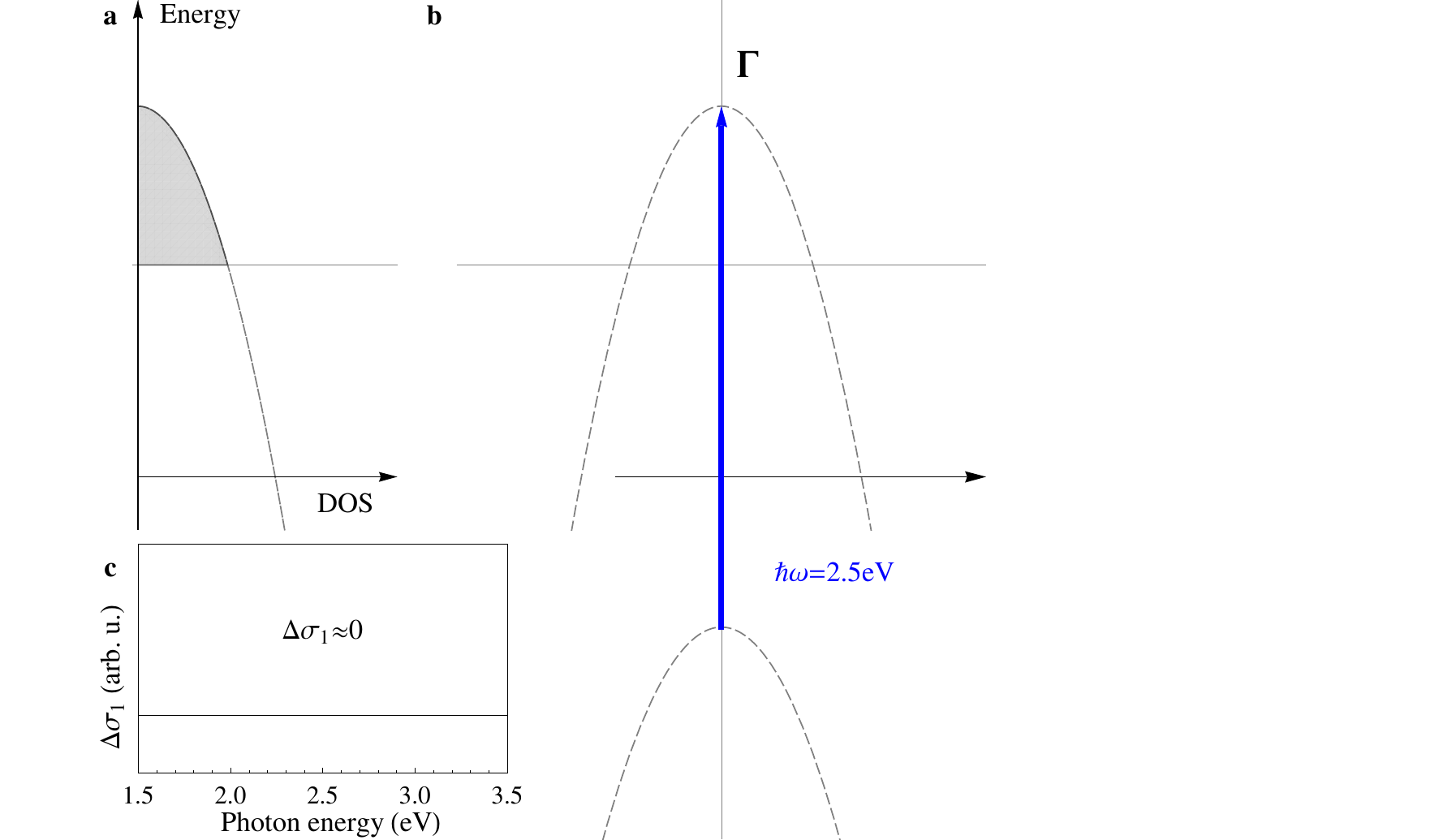}
\end{ocg}
\hskip0in
\vskip-4.26in
\begin{ocg}{scstatesingle}{2}{0}
\includegraphics[width=\textwidth]{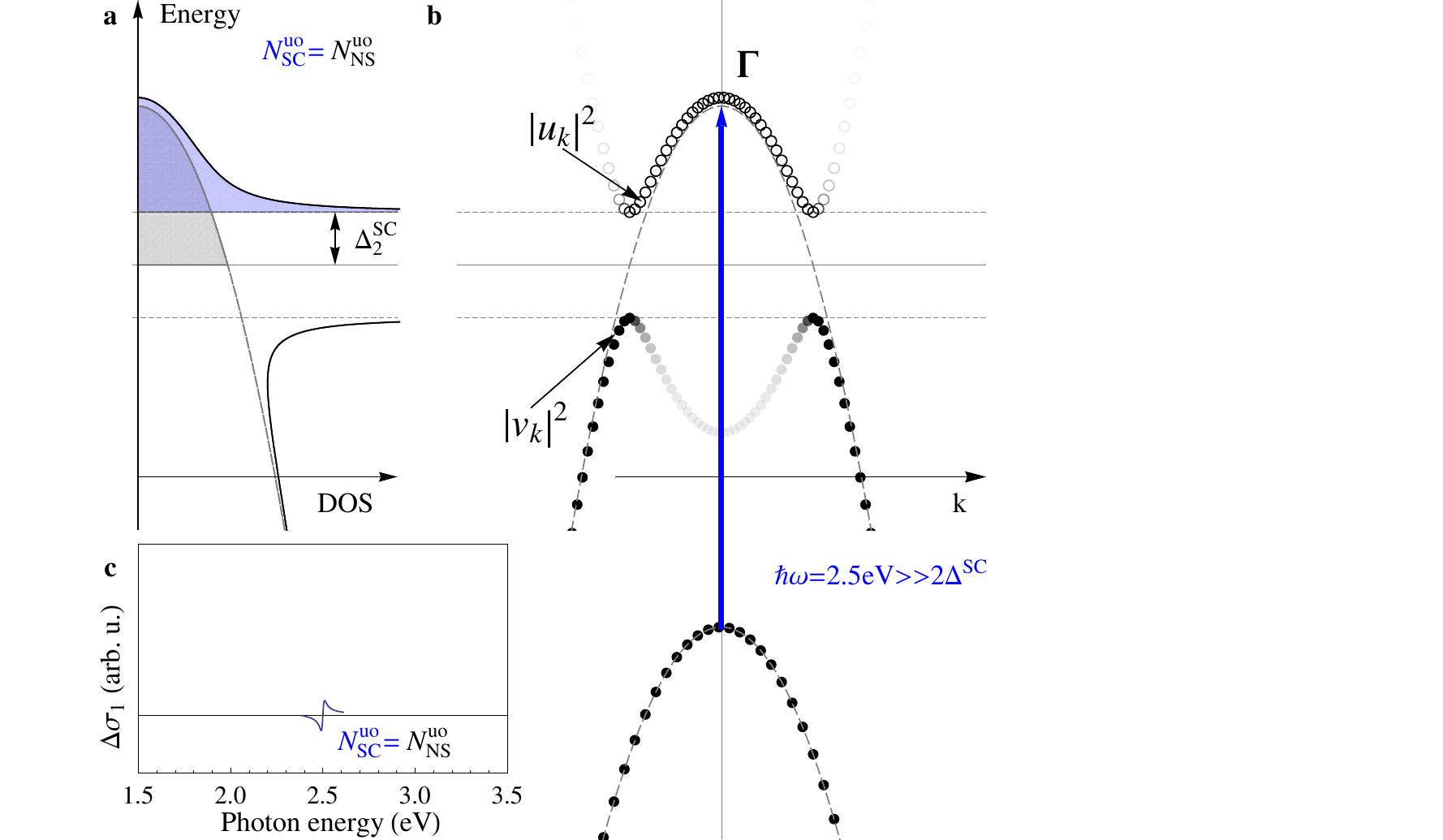}
\end{ocg}
\hskip0in
\vskip-4.26in
\begin{ocg}{scstatemulti}{3}{1}
\includegraphics[width=\textwidth]{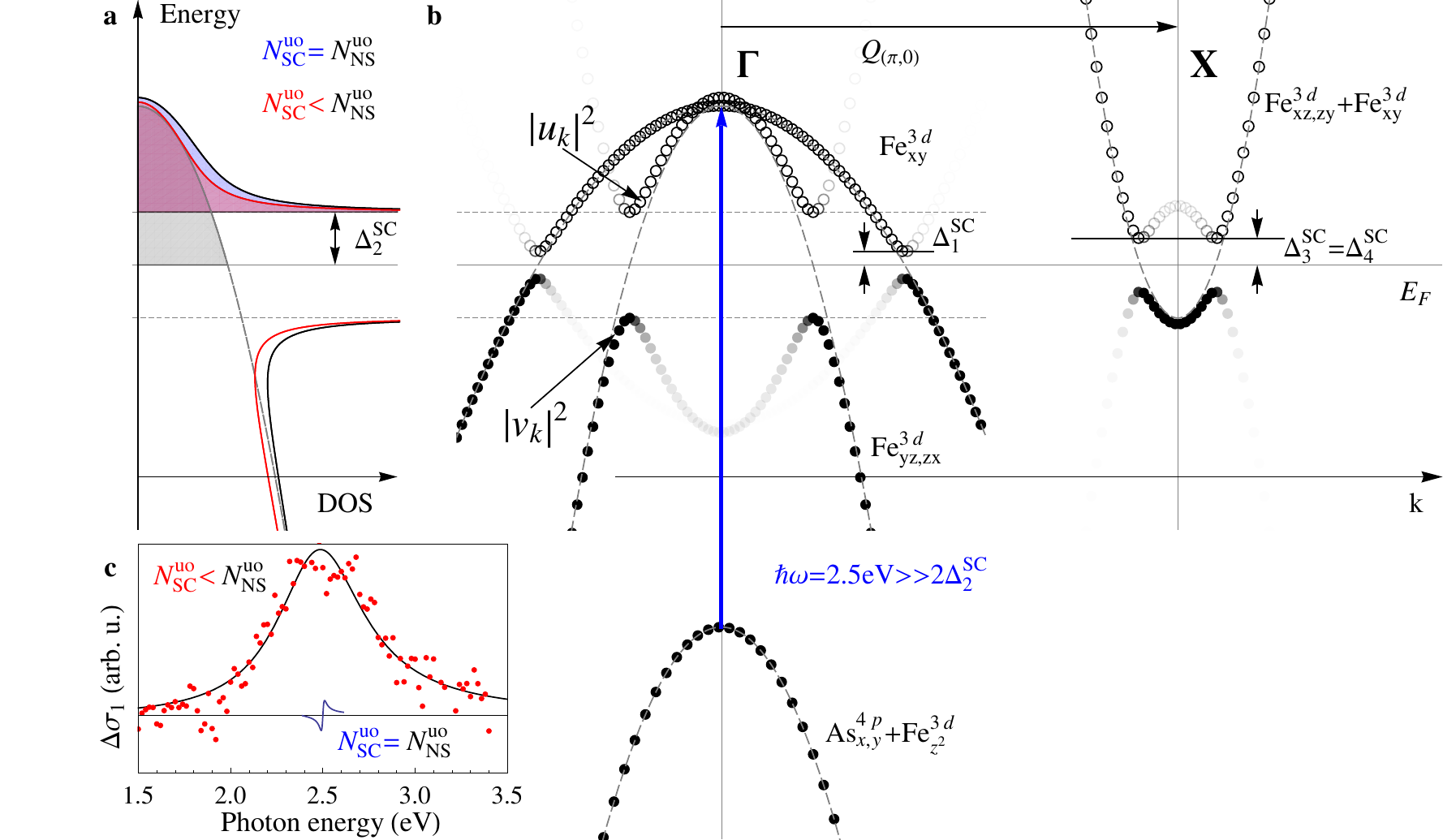}
\end{ocg}
\vskip-1.3in\hskip5in\begin{minipage}{1.5in}
\ToggleLayer{normalstate}{\framebox{\textbf{NORMAL STATE}}}\hfill\vskip0.25in
\ToggleLayer{scstatesingle}{\framebox{\textbf{SINGLE-BAND BCS}}}\hfill\vskip0.25in
\ToggleLayer{scstatemulti}{\framebox{\textbf{EXPERIMENT}}}\hfill
\end{minipage}
\vskip0.25in
\caption{\label{fig:bogolubov}(interactive: press buttons to choose) \textbf{a}~Density of states in the normal (gray line), conventional superconducting state (black line), and an unconventional state with a depletion of unoccupied states (red line). Filled areas of respective colors represent total number of unoccupied states. \textbf{b}~Schematic representation of the band structure of BKFA in the normal (dashed parabolas) and the superconducting state (filled and empty circles for occupied with probability $|v_{\mathbf{k}}^{ }|^2$ and unoccupied with probability $|u_{\mathbf{k}}^{ }|^2$ states with Bogolubov dispersion, respectively). The pair $(u_{\mathbf{k}}^{ },v_{\mathbf{k}}^{ })$ is unique for each separate band in the conventional multiband BCS approach~\cite{SPhysRev.108.1175,SPhysRevLett.3.552}. The orbitals of dominant contribution to each particular band are specified. \textbf{c}~(schematic) Difference spectra of the real part of optical conductivity between $40$ and $10\ \textrm{K}$.}
\end{figure}

One of the important and quite intuitive consequences of the Bogolubov-Valatin transformation~(\ref{eq:bogolubov}) is that the operators $\gamma_{\mathbf{k}\sigma}^{\mathstrut}$ and $c_{\mathbf{k}\sigma}^{\mathstrut}$ are connected via a unique, one-to-one relation. This immediately implies conservation of the total number of states in a given energy range, i.e.\begin{equation}N_{\mathrm{SC}}(E)dE=N_{\mathrm{NS}}(\xi)d\xi\label{eq:populationconservation},\end{equation}where $N_{\mathrm{SC}}(E)$ is the quasiparticle density of states (DOS) in the superconducting state, $N_{\mathrm{NS}}(\xi)$ is the normal-state electron DOS. This relation requires that the states within one $\Delta$ below the Fermi level be expelled to energies lower than $E_{\mathrm{F}}-\Delta$, while those within one $\Delta$ above the Fermi level to energies higher than $E_{\mathrm{F}}+\Delta$, as illustrated schematically in Fig.~\ref{fig:bogolubov}a. This process conserves the population above and below the Fermi level so that $N_{\mathrm{SC}}^{\mathrm{uo}}=N_{\mathrm{NS}}^{\mathrm{uo}}$ (blue area equal to the gray area in Fig.~\ref{fig:bogolubov}a). The exact analytical expression for the DOS of Bogolubov quasiparticles follows from equation~(\ref{eq:populationconservation}) bearing in mind that from the definition of the quasiparticle dispersion  $\xi_{\mathbf{k}}^{\mathstrut}=\sqrt{E_{\mathbf{k}}^{2}-\Delta^2}$:\begin{equation}N_{\mathrm{SC}}(E)=N_{\mathrm{NS}}(\xi)\frac{d\xi}{dE}=N_{\mathrm{NS}}(\xi(E))\frac{E_{\mathbf{k}}^{\mathstrut}}{\sqrt{E_{\mathbf{k}}^{2}-\Delta^2}}.\label{eq:scdos}\end{equation}%
This expression is plotted in Fig.~\ref{fig:bogolubov}a (black solid line) for the case of free-electron normal-state dispersion (gray dashed line). The square-root singularity in the quasiparticle DOS stems from the flattening of the normal-state dispersion in the vicinity of the Fermi surface, as shown in Fig.~\ref{fig:bogolubov}b (the inner hole dispersion corresponds to the DOS plotted in Fig.~\ref{fig:bogolubov}a). Occupied quasiparticle states are depicted as filled black circles, while the quasiparticle vacancies are shown as empty black circles. The fading of filled and empty black circles represents the occupation probabilities $(|u_{\mathbf{k}}^{\mathstrut}|^2,|v_{\mathbf{k}}^{\mathstrut}|^2)$ in equation~(\ref{eq:scprobabilities}). As it has already been mentioned above there exists a finite smearing of these probabilities even at $0\ \textrm{K}$. It leads to finite occupation of those regions of the Brillouin zone unoccupied in the normal state, the so-called \textit{backfolding} of the quasiparticle dispersion, clearly visible in Fig.~\ref{fig:bogolubov}b. As optical conductivity only probes averaged over wavevectors regions of the $\textbf{k}$-space it cannot resolve the result of the backfolding as opposed to angle-resolved photoemission spectroscopy (ARPES), where this effect has been reliably established~\cite{SPhysRevB.79.054517,SLeeVishik_backfolding_2007,SPhysRevB.53.R14737}. On the other hand the effect of smearing itself is incorporated in the density of states within the BCS formalism and is, therefore, included into our considerations. It is a single-band effect of superconductivity on the band structure and does not lead to population redistribution, i.e. the total number of unoccupied states below the transition is conserved $N^{\mathrm{uo}}_{\mathrm{SC}}=N^{\mathrm{uo}}_{\mathrm{NS}}$ (blue area is equal to the gray area in Fig.~\ref{fig:bogolubov}a). This can only lead to a small corrugation of an optical absorption band on the scale of one superconducting-gap energy superimposed on the overall broad feature without any modification of its spectral weight~\cite{SDobryakov1994309}. The experimentally observed suppression of an absorption band \textit{on the scale of its full width} rather than on the scale of superconductivity-induced modification of the dispersion necessarily requires population imbalance $N^{\mathrm{uo}}_{\mathrm{SC}}<N^{\mathrm{uo}}_{\mathrm{NS}}$ (red area is unequal to the gray area in Fig.~\ref{fig:bogolubov}a). It is unlikely that such an effect can be a consequence of a dynamic population balance between two or more bands at finite temperatures because the temperature dynamics of the suppression mimics that of the optical conductivity in the FIR region due to the opening of the superconducting gap, as shown in Fig.~\ref{fig:bkfatms}b.

To account for the multiband character of iron pnictides one may consider the multiband BCS theory~\cite{SPhysRevLett.3.552}. It is a straightforward generalization of the single-band BCS theory with the only complication that each separate band has its own gap, quasiparticle dispersion, and a pair $(u_{\mathbf{k}}^{\mathstrut},v_{\mathbf{k}}^{\mathstrut})$. However, in the framework of this multiband theory the quasiparticle operators do not involve normal-state particle operators and mix only the creation and annihilation operators from the same band (preferring our notation to that of ref.~\onlinecite{PhysRevLett.3.552}):
\begin{eqnarray}
c_{\mathbf{k}\uparrow}^{\mathstrut}&=&u_{\mathbf{k}}^{(c) *}e_{\mathbf{k}0}^{\mathstrut}+v_{\mathbf{k}}^{(c)}e_{\mathbf{k}1}^{\dagger},\nonumber\\
c_{-\mathbf{k}\downarrow}^{\mathstrut}&=&-v_{\mathbf{k}}^{(c) *}e_{\mathbf{k}0}^{\mathstrut}+u_{\mathbf{k}}^{(c)}e_{\mathbf{k}1}^{\dagger},\nonumber\\
d_{\mathbf{k}\uparrow}^{\mathstrut}&=&u_{\mathbf{k}}^{(d) *}f_{\mathbf{k}0}^{\mathstrut}+v_{\mathbf{k}}^{(d)}f_{\mathbf{k}1}^{\dagger},\nonumber\\
d_{-\mathbf{k}\downarrow}^{\mathstrut}&=&-v_{\mathbf{k}}^{(d) *}f_{\mathbf{k}0}^{\mathstrut}+u_{\mathbf{k}}^{(d)}f_{\mathbf{k}1}^{\dagger}\label{eq:bogolubovmulti},\end{eqnarray}where ($c_{\mathbf{k}\uparrow}^{\mathstrut},d_{\mathbf{k}\uparrow}^{\mathstrut}$) are normal-state particle operators and ($e_{\mathbf{k}\uparrow}^{\mathstrut},f_{\mathbf{k}\uparrow}^{\mathstrut}$) are the multiband counterparts of the operators $\gamma_{\mathbf{k}\uparrow}^{\mathstrut}$ in the single-band BCS theory. The coefficients $(u_{\mathbf{k}}^{(c,d)},v_{\mathbf{k}}^{(c,d)})$ certainly depend on the properties of both bands as well as on the interband coupling but the relations~(\ref{eq:bogolubovmulti}) are still unique one-to-one relations, which immediately implies that, however complicated the quasiparticle dispersions may be, for each separate band relation \begin{equation}N_{\mathrm{SC}}^{(i)}(E)dE^{(i)}=N_{\mathrm{NS}}^{(i)}(\xi)d\xi\label{eq:populationconservationmulti},\end{equation}holds, with $(i)$ running through all bands. As a consequence, just like in the single-band case, the total occupied/unoccupied population is conserved across the superconducting transition and thus only changes of interband optical conductivity on the scale of $2\Delta$ are expected. The population imbalance $N^{\mathrm{uo}}_{\mathrm{SC}}<N^{\mathrm{uo}}_{\mathrm{NS}}$ (red area in Fig.~\ref{fig:bogolubov}a smaller than the gray area) required to accommodate the experimentally observed suppression of an $2.5\ \textrm{eV}$ absorption band on its full width of about $1\ \textrm{eV}$ can come from redistribution of occupation of the different bands below $T_{\mathrm{c}}$. It requires a lowering of the material's chemical potential in the superconducting state and, therefore, an additional contribution to the condensation energy. However, the standard BCS theory and its generalization to the multiband case do not self-consistently take into account this effect: though predicting a lowering of a chemical potential as a consequence of a non-zero gain in the free energy of the system (condensation energy), they premise on equations with an essentially constant chemical potential. Consistent treatment of a variable chemical potential might render the Bogolubov-Valatin transformation~\ref{eq:bogolubovmulti} inappropriate in the multiband case and violate the population conservation of the occupied and unoccupied states within each band (as shown in Fig.~\ref{fig:bogolubov}a) - a fundamental consequence of the standard BCS theory. The resulting correction, small as it may be for conventional superconductors, in the presence of large Fe-As bond polarizability can lead to a large effect and potentially enhance superconductivity in iron pnictides. 
\subsection{Spectral weight and kinetic energy gain}
Increase of the condensation energy due to the lowering of the chemical potential of the system in the superconducting state explained in the previous chapter can be related to the SW of the suppressed optical band at $2.5\ \textrm{eV}$. Based on the current experimental evidence, it is impossible to confidently specify whether the liberated upon the superconducting transition SW contributes to that of the superconducting condensate. Due to the reliably determined background shift of the real part of $\Delta\varepsilon_1(\omega)$ in the visible spectral range (see Fig.~\ref{fig:bkfatms}a, bottom panel) it is certain, however, that this SW is redistributed at energies below $2\ \textrm{eV}$. One can easily estimate the order of magnitude of the free-energy gain resulting from this SW transfer assuming that it contributes to the intraband itenerant response~\cite{SPhysRevB.16.2437,SHirsch1992305}: $\Delta SW(\Omega)=(\pi e^2a^2/2\hbar^2V_u)\left<-E_{\rm{K}}\right>$, where $SW(\Omega)=\int_{0^+}^\Omega\sigma_1(\omega)d\omega$ is the in-plane spectral weight, $a$ is the in-plane lattice constant, $V_u$ - unit cell volume and $\left<-E_{\rm{K}}\right>=(1/N)\sum_{\vec k, \sigma}n_{\vec k, \sigma}\partial^2\varepsilon_{\vec k}/\partial k^2_\alpha$ is a measure of the system's kinetic energy per unit cell. In the case of the tight-binding nearest-neighbor approximation it is exactly equal to the kinetic energy of the charge carriers per unit cell and thus contributes directly to the condensation energy. This approximation might not hold for iron-pnictide superconductors but the above estimate shows that though small this additional spectral weight in the superconducting $\textrm{Ba}_{0.68}\textrm{K}_{0.32}\textrm{Fe}_2\textrm{As}_2$ is enough to account for the condensation energy in this compound.

\end{document}